\documentclass[12pt]{iopart}
\usepackage{graphicx}
\begin{document}

\title[Order in Graphene]{Pseudospin Order in Monolayer, Bilayer, and Double-Layer Graphene}

\author{A.H. MacDonald, Jeil Jung, and Fan Zhang}

\address{Department of Physics, University of Texas at Austin, Austin TX 78712 USA}
\ead{macd@physics.utexas.edu}
\begin{abstract}
Graphene is a gapless semiconductor in which conduction and valence band wavefunctions differ 
only in the phase difference between their projections onto the
two sublattices of the material's two-dimensional honeycomb crystal structure.  We explain why this
circumstance creates openings for broken symmetry states, including antiferromagnetic states in
monolayer and bilayer graphene and exciton condensates in double-layer graphene, that are
momentum space analogs of the real-space order common in systems with strong local interactions.
We discuss some similarities among, and some differences between, these three broken symmetry states.
\end{abstract}

\maketitle

\section{Introduction}

Graphene\cite{phystoday_review,rmp_review} is a two-dimensional crystal consisting entirely of carbon atoms.
Its honeycomb lattice is stabilized primarily by strong planar
$sp^{2}$ bonds, leaving one more weakly bonded $\pi$-orbital per carbon atom available
for metallic conduction.  Because of the two atoms per unit cell in a honeycomb lattice
the $\pi$ electrons form two bands, one of which is occupied in a neutral sheet.
Because the $\pi$-bands cross at the inequivalent honeycomb lattice Brillouin-zone
corners, {\em i.e.} at the $K$ and $K'$ BZ corner points, the gap between the occupied $\pi$ valence band and the
unoccupied $\pi$ conduction band vanishes.
Most graphene properties of interest to condensed-matter physicists,
for example transport and optical properties, directly involve only the $\pi$ electron
orbitals that are close to the band crossing points.  These are accurately described
over an energy interval several eV in width by a $\vec{k}\cdot\vec{p}$ Hamiltonian that
has the form of a massless Dirac equation:
\begin{equation}
\label{hband}
{\cal H}_{\rm band} = \hbar v_{0} \;  \sum_{\vec{k}, s',s} \;
c^{\dagger}_{\vec{k}, s'} (\vec{k} \cdot \vec{\sigma}_{s's})   c_{\vec{k},s}\,.
\label{eq:band}
\end{equation}
In Eq.(\ref{hband}) $v_{0}$ is the velocity of band electrons at the band-crossing (Dirac) point
which can be related to band Hamiltonian $\pi$-electron hopping amplitudes,   
and $\vec{\sigma}$ are Pauli matrices which act on sublattice labels $s(s')$.
Eq.(\ref{hband}) applies near the valley $K$ Dirac point; the corresponding equation for
valley $K'$ is obtained by letting $k_x \to - k_{x}$.  When the sublattice degree-of-freedom is viewed as
a pseudospin, we see from Eq.(\ref{hband}) that the band eigenstates are chiral.  
This is in the conduction band of valley $K$ pseudospin is parallel to the momentum, while 
in the valence band pseudospin is antiparallel to momentum.

Because only one of the two band states is occupied at each momentum, the many-body ground state
can be continuously deformed relative to the non-interacting ground state without breaking translational
symmetry simply by rotating the pseudospin direction at each momentum.  When interactions are neglected
the valence band is full, implying that the pseudospin direction is always opposite to the direction of momentum.
Since the interaction energy is minimized when all pseudospins are parallel, as we will discuss
explicitly below, there is tension between band energy minimization and interaction energy
minimization in monolayer graphene.  A similar tension arises in two other graphene based two-dimensional
electron systems -- graphene bilayers\cite{mccann_BLG} which consist of two Bernal stacked graphene layers and graphene double-layers\cite{min_exciton}
which consist of two layers separated by an insulating tunnel barrier.  In this article, we will discuss the
competition between interaction and band energies in all three systems using a common mean-field language which
enables us to highlight similarities and point out differences.
We will discuss the potential broken symmetry states using a mean-field-theory
in which the ground state is determined by minimizing
the total energy of single-Slater-determinant many-body states
varying the pseudospin direction at
each momentum $\vec{k}$.  By performing a stability analysis for the resulting energy 
functional we conclude that broken symmetries 
states can occur in monolayer graphene if interactions are sufficiently strong, and that they occur in bilayer and
double-layer graphene for interactions of any strength.  Our mean-field theory treats the
gapped\cite{min_MF,zhang_RG,levitov_MF} and nematic states\cite{vafek_RG,falko_RG} 
that have been discussed for bilayer graphene on an equal footing.

A parallel can be drawn between the $\pi$-orbital states in 
these three graphene systems and the electronic states of Mott insulators.  In Mott insulators 
strong interactions project most of the many-electron wavefunction onto a subspace with 
one atom per unit cell.  The only degree-of-freedom that is available in this subspace is the spin state at 
each lattice position.  Unless interactions between spin orientations on 
different lattice sites are frustrated, the ground state is normally close to the classical ground state 
in which spin-orientations are fixed on each site and chosen to minimize the expectation 
value of the effective spin Hamiltonian.  Quantum fluctuations of spin orientations play only a  
quantitative role.  For the graphene states discussed here the pseudospin energy 
function that is fixed by energy minimization plays a 
role similar to the spin distribution in an insulator, but is a function of momentum rather than position.  
The graphene system band Hamiltonians, which differ essentially in the three cases,
act like momentum-dependent pseudospin fields.  Because the band Hamiltonians 
reduce symmetries, energy minimization does not 
in all cases imply broken symmetry ground states.  Because of the absence of a 
gap in the graphene case, important quantum fluctuations
occur in both pseudospin orientations and in the 
occupation numbers of momentum states, but we expect that their role is also only quantitative and 
we do not discuss them at length in this paper.  
The broken symmetry states which can occur in these graphene systems are unusual from several points of view.   
For a given spin and valley, the ordered states of both 
monolayer and bilayer graphene have large quasiparticle Berry curvature and spontaneous quantized
anomalous Hall effects\cite{xiao,levitov_QAH,zhang_SQH,jung_MF,zhang_ZM}.
The potential broken symmetry states of double-layer graphene have spontaneous interlayer phase coherence,
which leads to the suite of phenomena connected with counter-flow superfluidity \cite{jpe_Nature, su} 
when the layers are separately contacted. 

Our paper is organized as follows.  In Section $2$ we formulate the mean-field theory in the three
graphene-based two-dimensional electron systems in terms of energy minimization with
respect to $\vec{k}$-dependent pseudospin orientation.  In Sections $3$, $4$, and $5$ we discuss
a stability analysis of the energy functional at the band state configuration
for monolayer, bilayer, and double-layer systems respectively and comment on the
character of the broken-symmetry states that result.  Finally in Section $6$ we
point to important similarities and
differences between the three cases.

\section{Pseudospins in Monolayer, Bilayer, and Double-Layer Graphene}

We consider a variational single-Slater-determinant wavefunction with a single pseudospin state occupied
at each momentum $\vec{k}$, and minimize the expectation value of the Hamiltonian
with respect to the pseudospin orientations $\hat{n}_{\vec{k}}$.  In this approximation the
four spin-valley flavors interact only through their contribution to the electrostatic Hartree energy.
Since we will consider only states that are locally electrically neutral, we will
ignore the Hartree energy for the moment and return to it later when we 
discuss the role of spin-valley flavors.

\begin{figure}[hbp]
\label{blochsphere}
\begin{tabular}{cc}
\quad \quad \quad \quad \quad \quad& \includegraphics[width=12cm,angle=0]{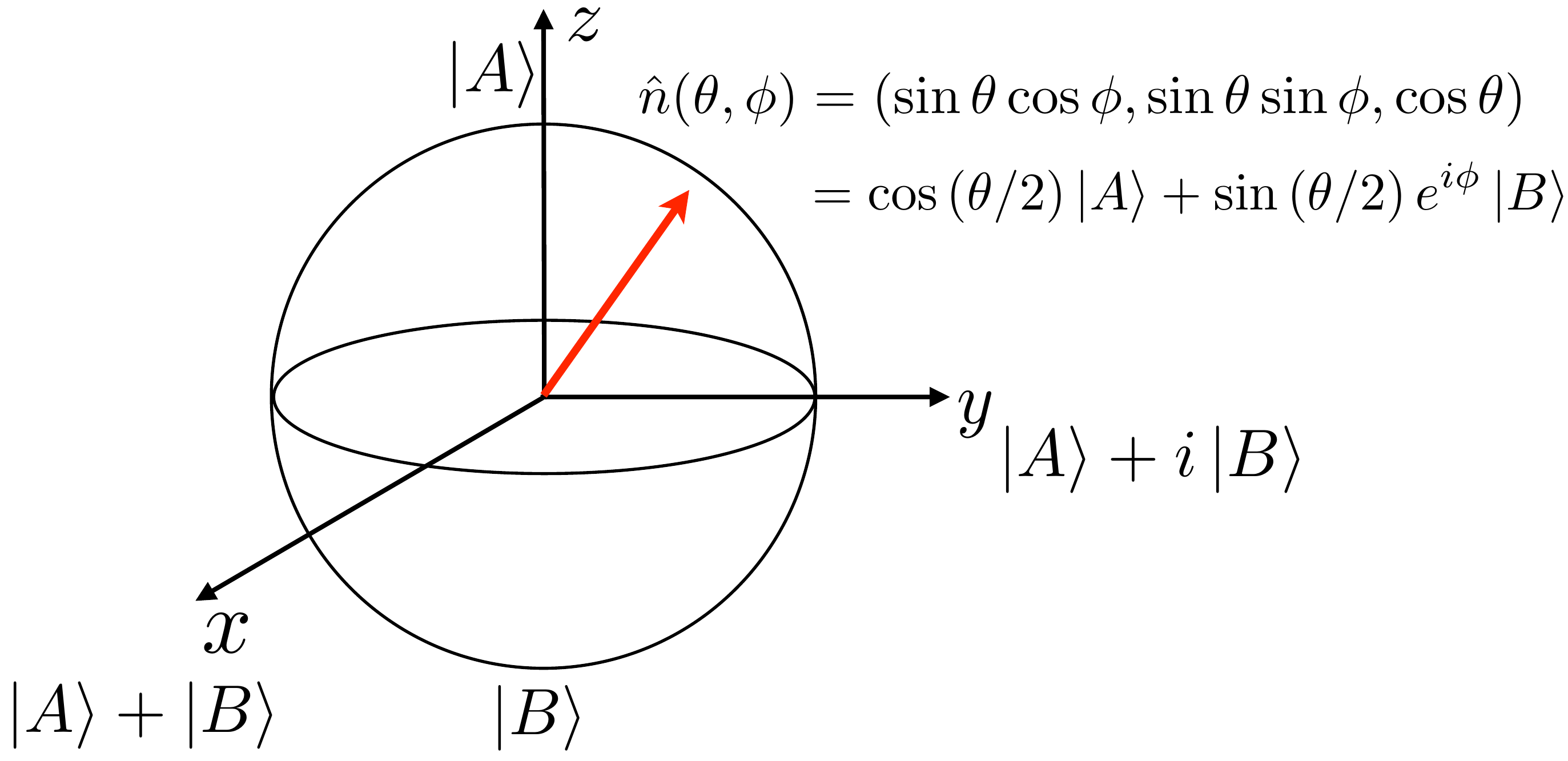}   \\
\end{tabular}
\caption{Representation of a pseudospin state by a unit vector 
$\hat{r}(\theta,\phi)$.  Pseudospins directed along the $\pm z$ direction 
represent the states labeled $\left| A \right>$ and $\left| B \right>$.
For monolayer graphene these states are confined to alternate sublattices,
while for bilayer and double layer graphene they are confined to alternate layers. 
General pseudospin states are coherent linear combinations of 
$\left| A \right>$ and $\left| B \right>$, with the azimuthal pseudospin 
angle $\phi$ specifying the phase difference between $\left| A \right>$ and $\left| B \right>$
amplitudes and the polar pseudospin angle specifying the $\left| A \right>$-$\left| B \right>$ 
polarization.   A state with $\theta = \pi/2$ has equal $\left| A \right>$ and $\left| B \right>$ weights,
whereas the states with $\theta=0$ and $\theta=\pi$ equal $\left| A \right>$ and $\left| B \right>$ 
respectively.
}
\end{figure}

It will be convenient to express the band Hamiltonian as an effective
magnetic field that acts on the pseudospin degree-of-freedom by writing
\begin{equation}
\label{hbandpsspin}
{\cal H}_{\rm band} =  \;  \sum_{\vec{k}, s',s} \;
c^{\dagger}_{\vec{k}, s'} (\vec{h}_{\vec{k}} \cdot \vec{\sigma}_{s's})   c_{\vec{k},s}\,.
\label{eq:band}
\end{equation}
In this language the three cases we discuss are distinguished by their pseudospin effective
magnetic fields:
\begin{eqnarray}
\vec{h}_{\vec{k}}^{\rm (ML)} = \hbar v_{0} k \; [ \cos(\phi_{\vec{k}})\, \hat{x} + \sin(\phi_{\vec{k}})\, \hat{y} ]\,,  \nonumber \\
\vec{h}_{\vec{k}}^{\rm (BL)} = -(\hbar^2 k^2 /2m^*) \; [ \cos(2\phi_{\vec{k}})\, \hat{x} + \sin(2\phi_{\vec{k}})\, \hat{y} ]\,,  \nonumber \\
\vec{h}_{\vec{k}}^{\rm (DL)} = \hbar v_{0} (k-k_{F})\, \hat{z}\,,
\end{eqnarray}
for monolayer, bilayer, and double-layer cases respectively.  Here $\phi_{\vec{k}}$ is the
angular orientation of the two-dimensional  $\vec{k}$   
momentum and $k$ is its magnitude.
Note that the band Hamiltonians are off-diagonal
in pseudospin index in the monolayer and bilayer cases, and diagonal in the double-layer case.
The pseudospin labels in Eq.(\ref{eq:band}) refer to the sublattice index
 in the monolayer graphene case and to layer index in the bilayer and double-layer cases.  The form we have chosen for the bilayer Hamiltonian applies only at energies smaller than the
interlayer hopping energy $\gamma_1$ and is due\cite{mccann_BLG} to virtual hopping between the two low-energy
bilayer $\pi$-orbital sites, which are located in different layers, via two higher-energy $\pi$-orbital sites that 
are not explicitly retained.
Because the pseudospin labels refer to position in all three cases, the electron-electron
interaction Hamiltonian is diagonal in pseudospin at each vertex in all three cases:
\begin{equation}
\label{hint}
{\cal H}_{\rm int} =  \frac{1}{2A}   \sum_{\vec{k}, \vec{p},\vec{q}} \sum_{s,s'} \;
c^{\dagger}_{\vec{k}+\vec{q}, s} c^{\dagger}_{\vec{p}-\vec{q}, s'}   c_{\vec{p},s'} c_{\vec{k},s} \,
\left[ V^+(\vec{q}) + V^-(\vec{q}) \sigma^{z}_{ss} \sigma^{z}_{s's'}\right]
\end{equation}
where $V^{\pm}(\vec{q})=(V_{\rm S} \pm V_{\rm D})/2$ and $V_{\rm S,D}$ are the momentum-space
interactions between electrons on the same and different sublattices 
(or layers) 
respectively.
In the monolayer case the interactions are pseudospin independent ($V_{\rm S}=V_{\rm D}$), whereas in the
bilayer and double-layer cases they are pseudospin dependent because interlayer interactions
are weaker than intralayer interactions.

The dependence of the band energy on the momentum-dependent pseudospin configuration is easy to evaluate.
From the form of the pseudospinor for a state aligned in a particular pseudospin direction (see Fig.[1]),
\begin{equation}
\label{coherentstate}
|\hat{n}\rangle =
\left(
\begin{array}{c}
 \cos\frac{\theta}{2} \\
 \sin\frac{\theta}{2}\, e^{\rm i\phi}
\end{array}
\right)\,,
\end{equation}
it is easy to show that
\begin{equation}
\langle \hat{n} | \vec{\sigma} | \hat{n}\rangle = \hat{n}\,,
\end{equation}
and that
\begin{equation}
|\hat{n}\rangle \langle \hat{n}| = \frac{1  + \vec{\sigma} \cdot \hat{n}}{2}.
\end{equation}
In Eq.(\ref{coherentstate}), $\theta$ and $\phi$ are the polar and azimuthal angles for the
pseudospin orientation, {\em i.e.} $\hat{n} = \left(\sin\theta \cos\phi, \sin\theta \sin\phi, \cos\theta\right)$.
For the case of pseudospin-dependent interactions the following identity is useful:
\begin{equation}
\sigma^{z} |\hat{n}\rangle \langle \hat{n}| \sigma^{z} = \frac{1  - \sigma^x n^x - \sigma^y n^y + \sigma^z n^z }{2}.
\end{equation}
Taking the expectation values of the band and interaction Hamiltonians we find that
\begin{equation}
\label{eband}
E_{\rm band}[\hat{n}_{\vec{k}}]  =  \;  \sum_{\vec{k}} \;
\vec{h}_{\vec{k}} \cdot \hat{n}_{\vec{k}}\,,
\end{equation}
and that 
\begin{equation}
\label{eint}
E_{\rm int}[\hat{n}_{\vec{k}}] = \frac{-1}{4A}\sum_{\vec{k}, \vec{p}}
\left[ \big(1+n^z_{\vec{k}} n^z_{\vec{p}}\big)  \; V_{\rm S}(\vec{k}-\vec{p}) 
+ \big(n^x_{\vec{k}} n^x_{\vec{p}}+ n^y_{\vec{k}} n^y_{\vec{p}}\big) \; 
V_{\rm D}(\vec{k}-\vec{p}) \right].
\end{equation}
The interaction energy is an exchange contribution which sets the 
momentum transfer $\vec{q}$ in Eq.(\ref{hint}) to
$\vec{p}-\vec{k}$.
As explicitly shown in Eq.(~\ref{eint}), the exchange energy contribution from any
pair of momentum is lowered when their pseudospins are made more parallel. 
The band energy, on the other hand,
is minimized when the pseudospin direction is opposite to the direction of 
$\vec{h}_{\vec{k}}$ at every $\vec{k}$ and hence strongly pseudospin dependent.  
In the following sections we will discuss the pseudospin orientation function 
which minimizes the total energy $E_{\rm tot}=E_{\rm band}+E_{\rm int}$ for 
monolayer, bilayer, and double layer cases.

\section{Spin-Density-Wave States in Monolayer Graphene}
\begin{figure}[hbp]
\label{monolayerfig}
\begin{tabular}{cc}
\quad \quad \quad \quad \quad \quad& \includegraphics[width=12cm,angle=0]{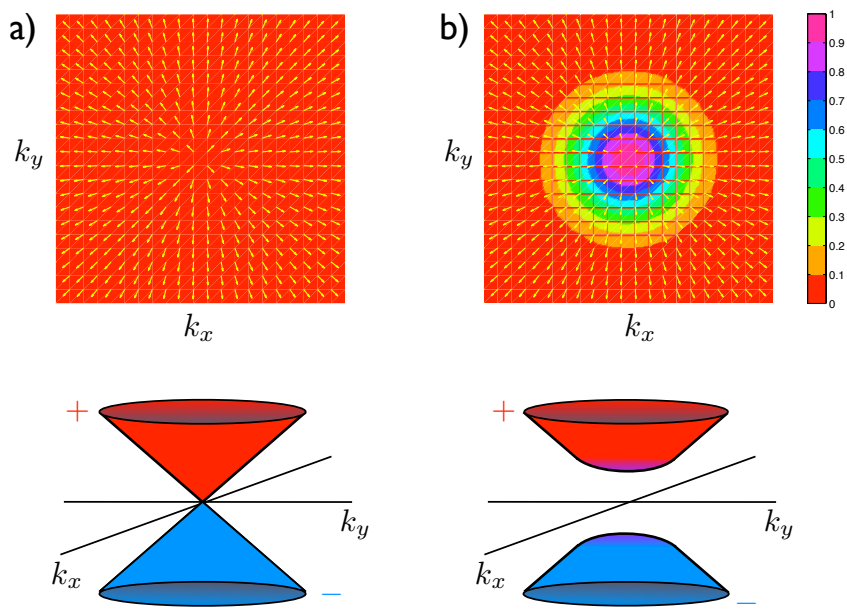}    \\
\end{tabular}
\caption{Pseudospin orientations and quasiparticle spectra near the Dirac point for gapless and 
gapped monolayer graphene.
The upper row illustrates conduction band pseudospin orientations for 
gapless (a-left) and gapped (b-right) cases; the 
valence band pseudospins are opposite in direction at each momentum. 
The arrows indicate the magnitude and direction of the $\hat{x}-\hat{y}$ plane 
pseudospin projection while the color indicates the $\hat{z}$ direction pseudospin projection 
which is zero in the gapless unbroken symmetry state.   
In monolayer graphene pseudospins rotate around the $\hat{z}$ axis by $2 \pi$
when a point in momentum space encloses the Dirac point $\vec{k}=0$.
In the broken symmetry state the self-energy contribution to the 
pseudospin field has a component in the $\hat{z}$ direction which 
does not vanishes for $k \to 0$, opening up a gap in the quasiparticle spectrum.  
}
\end{figure}

We find it useful to expand the total energy functional to leading order around its band theory value
$\hat{n}_{{\rm b}\,\vec{k}}=-\hat{h}_{\vec{k}}$.
This consideration follows similar lines in the three cases of
interest.  To leading order we can preserve normalization by writing
\begin{equation}
\hat{n}_{\vec{k}}=-\hat{h}_{\vec{k}}\,\left(1-\frac{1}{2}|\vec{\delta}_{\vec{k}}|^2\right)
+ \vec{\delta}_{\vec{k}}
\end{equation}
where $\vec{\delta}_{\vec{k}}$ is a two-dimensional vector perpendicular to $\hat{h}_{\vec{k}}$.
For monolayer graphene the band pseudospin orientation
$\hat{n}_{{\rm b}\,\vec{k}} = - \hat{k} = - \left(\cos\phi_{\vec{k}},\sin\phi_{\vec{k}},0\right)$ and
$\delta$ has $\hat{z}$ and azimuthal ($\hat{\phi}$) components along $\hat{z}=(0,0,1)$ and
$\hat{\phi}_{\vec{k}} = \hat{z} \times \hat{k} = \left(\sin\phi_{\vec{k}},-\cos\phi_{\vec{k}},0\right)$ directions respectively.
We find that
\begin{equation}
E=E_{0} + \frac{1}{4A}  \sum_{\vec{k}, \vec{p}} \sum_{\alpha,\beta}  \;  \delta^{\alpha}_{\vec{k}} \,
 K_{\vec{k},\vec{p}}^{\alpha,\beta} \, \delta^{\beta}_{\vec{p}}
\end{equation}
where
\begin{equation}
K_{\vec{k},\vec{p}}^{\alpha,\beta}  =   \delta_{\alpha,\beta} \big[  2A (h_{\vec{k}}+\Sigma_{\vec{k}}) \delta_{\vec{k},\vec{p}}
- V_{\rm D}(\vec{k}-\vec{p}) \, \hat{\phi}_{\vec{k}} \cdot \hat{\phi}_{\vec{p}} \; \delta_{\alpha,\phi} - V_{\rm S}(\vec{k}-\vec{p}) \, \delta_{\alpha,z} \big]
\end{equation}
where
\begin{equation}
\Sigma_{\vec{k}} = -\frac{\partial E_{\rm tot}[\hat{n}_{\vec{k}}]}{\partial \hat{n}_{\vec{k}}} \cdot \hat{h}_{\vec{k}}.
\end{equation}

The quantity $\Sigma_{\vec{k}}$ in the above equations, the change in interaction energy associated with 
switching a single band state from valence to conduction band pseudospin orientations, can be identified as the exchange contribution
to the self-energy of the band state.  $\Sigma_{\vec{k}}$ adds to the energy difference between
conduction and valence band states because of the energy cost of reversing 
pseudospin orientation at a single-momenta, keeping all other pseudospins fixed.
For the monolayer graphene case we can let $V_{\rm D} \to V_{\rm S} \to V$ so that
\begin{equation}
\Sigma_{\vec{k}} =  \frac{1}{2A} \sum_{\vec{p}} V(\vec{p}-\vec{k}) \; \hat{n}_{{\rm b}\,\vec{p}} \cdot \hat{n}_{{\rm b}\,\vec{k}},
\end{equation}
which vanishes for a $\delta$-function interaction model because of the angular average over the direction of $\vec{p}$,
but grows with $|\vec{k}|$ for the realistic Coulomb interaction case:
\begin{equation}
\Sigma_{\vec{k}} \simeq   \frac{\alpha}{4} \; \hbar v_{0} k \,  \ln(\Lambda/k)
\end{equation}
where $\alpha = e^2/\epsilon \hbar v_{0}$ is graphene's fine structure constant, 
$\Lambda \sim 1/a$ is the Dirac model ultraviolet cutoff where $a$ is graphene's lattice constant, and $\epsilon$ is 
the graphene sheet's effective dielectric constant which depends on the substrate used to support the sheet.
Full Brillouin zone Hartree-Fock theory calculations\cite{mono_hf} suggest that 
the most appropriate value for  $\Lambda$ is $\sim 30/a$.
The self-energy term captures the physics of a theoretically anticiapted 
\cite{mono_hf,abrikosov, logdivref, logdivref1, logdivref2}   
logarithmic interaction enhancement of the energy difference between
conduction and valence band quasiparticles, that has now been confirmed experimentally\cite{geim}.
This effect that is normally described in terms of
an interaction enhanced quasiparticle velocity at momenta near the Dirac points.

Depending on the dielectric environment of a monolayer graphene sheet, the fine structure constant
value can vary between $\alpha \sim 0.5$ and $\alpha \sim 2$\cite{fuhrer}.
It is sometimes claimed that perturbative and mean-field treatment of
electron-electron interaction effects, like the one discussed here,
cannot be trusted because the coupling constant is not small.   
A more reliable way to judge the adequacy of these
approximations is to compare with experiments.  
In the case of monolayer graphene application of this type of criteria 
usually argues for the opposite conclusion, namely that mean-field theory is
reliable, provided that the electron-electron interaction is properly screened when carriers are present.
The approximation in which electronic self-energies are approximated at leading order
in dynamically screened Coulomb interactions, variously known as the random phase
approximation (RPA) or the GW approximation, 
agrees very well with, in particular, photoemission experiments\cite{rotenberg}
in both neutral and charged monolayer graphene.  For neutral graphene the inverse
quasiparticle lifetime\cite{louie_prl,polini_prb} in neutral graphene is $\sim 10\%$ of the quasiparticle 
energy at typical values of $\alpha$,
and this ratio may provide a better characterization of interaction strength.

Although some details of graphene interaction physics are sensitive to the long range of 
Coulomb interactions\cite{mono_hf,honerkamp,herbut,wehling},
most importantly perhaps the velocity enhancement mentioned above,  
we are able to make a number of valuable points by considering a short-range 
interaction model in which the momentum
dependence of $V_{\rm S}$ and $V_{\rm D}$ are neglected.  Note that because we are neglecting
inter-valley scattering we are still imagining that the interaction range is large compared to the atomic length
scale, and that there is therefore no direct relationship between this approximation and using a lattice Hubbard model
\cite{fujita,afqmc1,afqmc2,meng}.
Setting $V_{\rm S} = V_{\rm D} \to U$ we find that $\Sigma$ vanishes, that
\begin{equation}
K_{\vec{k},\vec{p}}^{\phi,\phi}  \to     2A \hbar v_{0} k  \delta_{\vec{k},\vec{p}}
- U  \hat{\phi}_{\vec{k}} \cdot \hat{\phi}_{\vec{p}},
\end{equation}
and that
\begin{equation}
K_{\vec{k},\vec{p}}^{zz}  \to     2A \hbar v_{0} k \delta_{\vec{k},\vec{p}}- U \,.
\end{equation}
In-plane and out-of-plane pseudospin reorientation 
instabilities are indicated by vanishing eigenvalues for $K_{\vec{k},\vec{p}}^{\phi,\phi}$ and
$K_{\vec{k},\vec{p}}^{zz}$ respectively.  We search for a zero eigenvalue by solving
\begin{equation}
\frac{1}{A} \sum_{\vec{p}} \, K_{\vec{k},\vec{p}}^{\alpha,\alpha} \delta^{\alpha}_{\vec{p}}  \equiv 0.
\end{equation}
These homogeneous equations are solved by setting 
$\delta^{\phi}_{\vec{p}} \to C \cos(\phi_{\vec{p}}-\chi)/\hbar v_{0} k$
and $\delta^{z}_{\vec{p}} \to C/\hbar v_{0} k$ where $C$ is an arbitrary 
constant and $\chi$ an arbitrary angle.  We
obtain the conditions
\begin{equation}
2  \cos(\phi_{\vec{k}}-\chi)  = U  \int \frac{d\vec{p}}{(2\pi)^2}\,\frac{\cos(\phi_{\vec{p}}-\phi_{\vec{k}})\cos(\phi_{\vec{p}}-\chi)}{\hbar v_{0} p}
\end{equation}
for $K^{\phi,\phi}$ and
\begin{equation}
2 = U \; \int  \frac{d\vec{p}}{(2\pi)^2} \;  \frac{1}{\hbar v_{0} p}
\end{equation}
for $K^{z,z}$.  Converting the integral over momentum $\vec{p}$ into an integral over the
energy of quasiparticles with energy $\hbar v_{0} p$, it follows that instability occurs in 
$K^{\phi,\phi}$ when $U \nu^*  > 4$, whereas for
$K^{z,z}$ it occurs when $U \nu^* > 2$.  In these equations $\nu^* = W/(2\pi\hbar^2 v_{0}^2)$
is the Dirac-model density-of-states at the model's ultraviolet energy 
cutoff scale $W \sim \hbar v_{0}/a $; the integrand of the energy
integral is constant because of a cancellation 
between the $\hbar v_{0} p$ factor in the denominator and
the quasiparticle density-of-states, which is proportional to energy.  Interactions are less
effective in reducing the energy cost of in-plane $\phi$ distortions than out-of-plane 
$z$ distortions because of the angle-dependence of the
band state pseudospins.

There are two important points to make about these stability criteria results.  First of all, we see that if interactions
are strong, pseudospins are more likely to tilt toward the $\pm\hat{z}$ directions, rather than to alter their orientations in the
$\hat{x}-\hat{y}$ plane.  The state produced by this pseudospin distortion has a higher electron density on one
sublattice than the other,  hence it is a density-wave state when looked at from a microscopic point of view.
In ignoring the Hartree energy, we have implicitly assumed that the signs of the density-waves are opposite for opposite valleys or for
opposite spins, with the latter possibility being more likely as discuss later. 
The expected state is therefore a
spin-density-wave state rather than a charge-density-wave state.  If the instability involved distortion of the
in-plane pseudospin direction, rather than tilting out of the plane, it would yield a state with spontaneous anisotropy,
characterized by the angle $\chi$, in which the magnitude of the quasiparticle self-energy depends on momentum direction.
The mean-field-theory instability analysis therefore
suggests that density-wave instabilities occur before anisotropy instabilities. 
Secondly, we note that the instability criterion involves $\nu^*$, the density-of-states at the band width energy scale.
We can conclude from this observation that the presence or absence of a broken symmetry state depends on
atomic length scale physics beyond that captured by the $\vec{k}\cdot\vec{p}$ Dirac model.
\cite{jung_MF, mono_bfield, mono_hf}

When solved with an unscreened Coulomb interaction\cite{min_MF}, mean-field theory predicts
that the instability occurs in monolayer graphene at $\alpha \sim 1.3$; 
screening and other higher-order corrections will shift\cite{gonzales} the
instability - likely to larger values of $\alpha$ beyond those that can be reached in monolayer graphene even when
it is suspended so that interactions are not reduced by dielectric screening.  Degrees of freedom not included in the
$\pi$-band only model of graphene, and portions of the Brillouin-zone far from the the corner Dirac points are also likely
to play an important role.  The instability can also be ruled out experimentally with nearly complete confidence because
pseudospin orientations in the $\hat{z}$ direction open gaps in the quasiparticle spectrum.  It can be established
experimentally the gaps, if present, cannot be larger than $\sim 10^{-4}$ eV compared to the natural
energy scale of graphene which is $\sim 10$ eV.  \cite{bolotin}

\begin{figure}[hbp]
\label{bilayerfig}
\begin{centering}
\begin{tabular}{cc}
\quad \quad \quad \quad \quad \quad& \includegraphics[width=12cm,angle=0]{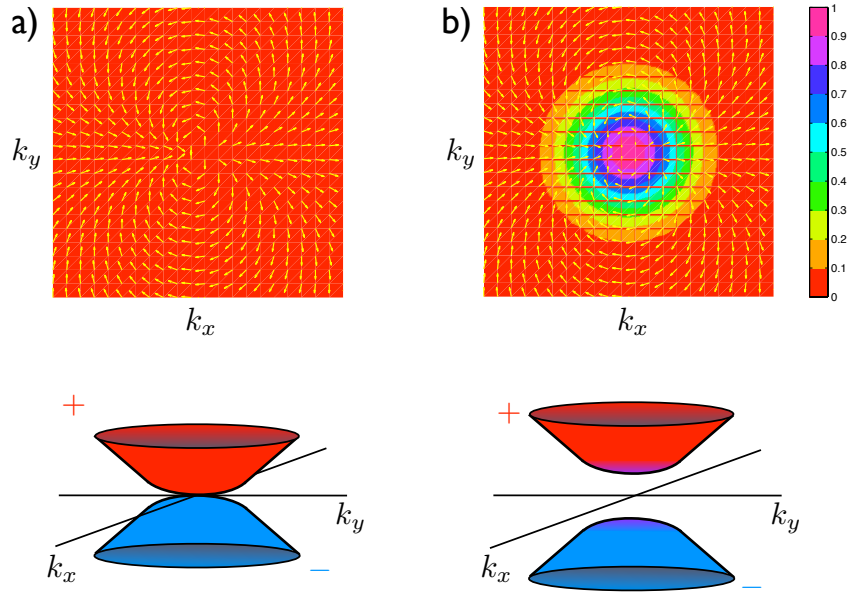}    \\
\end{tabular}
\end{centering}
\caption{
Pseudospin orientations and quasiparticle dispersions near the Dirac point in bilayer graphene.
The upper row illustrates the conduction band pseudospin orientations; the 
valence band pseudospin orientation is opposite at each momentum. 
In bilayer graphene pseudospins rotate by an angle of $4 \pi$
around the $\hat{z}$ direction when a point in momentum space encloses the 
$\vec{k}=0$ Dirac point.
The figures (a) and (b) represent respectively gapless and gapped broken symmetry states. 
As in the monolayer case  all pseudospins are in the $xy$ plane in the ungapped state.
In the gapful case the $\pm \hat{z}$  out of plane pseudospin components are non-zero 
and indicate 
sublattice polarization. In the low energy model for bilayers the sublattice index labels layer.} 
\end{figure}

\section{Antiferromagnetic States in Bilayer Graphene}

Several consequential distinctions can be drawn between the monolayer and bilayer cases.  First of all the two-band
$\vec{k}\cdot\vec{p}$ for bilayer model provides a good description only at energies that are small compared to the interlayer
tunneling energy $\gamma_1 \sim 0.4$ eV, whereas the corresponding monolayer two-band model
applies up to energies $\sim 2$ eV, much closer to the full band width.   Secondly, both the magnitude and
direction of pseudospin effective magnetic fields have different behaviors since $\vec{h}_{\vec{k}}^{\rm (BL)}$ varies quadratically rather than
linearly with $k$ and has an orientation angle that is twice the momentum orientation angle.  The stability analysis for bilayers
parallels that for monolayers precisely once these differences are recognized.  For bilayers the band and
in-plane-distortion pseudospin directions are  $\hat{n}_{{\rm b}\,\vec{k}} = \left(\cos(2\phi_{\vec{k}}),\sin(2\phi_{\vec{k}}),0 \right)$
and $\hat{\phi}_{\vec{k}} = \hat{n}_{{\rm b}\,\vec{k}} \times \hat{z}  = \left(\sin(2\phi_{\vec{k}}),-\cos(2\phi_{\vec{k}}),0 \right)$.
The more rapid variation of pseudospin direction with momentum direction eliminates the logarithmic divergence of
the velocity enhancement at small $k$ found in the monolayer Coulomb interaction case,
but still\cite{logdivref2} leaves a substantial interaction-induced velocity enhancement.  
For the short-range interaction model, the stability matrices for $\phi$ and $z$ distortions are
\begin{equation}
K_{\vec{k},\vec{p}}^{\phi,\phi}  \to     2A \, \frac{\hbar^2k^2}{2m^*}  \, \delta_{\vec{k},\vec{p}}
- U  \hat{\phi}_{\vec{k}} \cdot \hat{\phi}_{\vec{p}},
\end{equation}
and
\begin{equation}
K_{\vec{k},\vec{p}}^{zz}  \to     2A \, \frac{\hbar^2k^2}{2m^*} \, \delta_{\vec{k},\vec{p}}
- U
\end{equation}
where $\hat{\phi}_{\vec{k}} \cdot \hat{\phi}_{\vec{p}}$ now equals $\cos(2\phi_{\vec{p}}-2\phi_{\vec{k}})$.
Setting $\delta^{\phi}_{\vec{p}} \to C \cos(2\phi_{\vec{p}}-\chi)/(\hbar^2p^2/2m^*)$
and $\delta^{z}_{\vec{p}} \to C/(\hbar^2p^2/2m^*)$ we can solve for the interaction strength at
which the smallest eigenvalue approaches zero.  As in the monolayer case we find that the
$\pm\hat{z}$ distortions, which in this case correspond to moving charge between layers, occur at weaker
interaction strengths.  The instability criteria\cite{zhang_RG} in the bilayer case are $U \nu^{\rm BL} \ln(W/E_{F})   > 4$ for $K^{\phi,\phi}$ and
$U \nu^{\rm BL}  \ln(W/E_{F}) > 2$ for $K^{z,z}$.  In this case $\nu^{\rm BL}=m^*/(2\pi\hbar^2)$ is the energy-independent band electron density-of-states.  The $1/E$ quasiparticle energy factor is not canceled by an increasing
density-of-states, as it was in the monolayer case, and the interaction contribution to the stability eigenvalue
integral has a logarithmic infrared divergence which we have cut off by assuming that conduction band
states up to energy $E_{F}$ have been occupied, Pauli-blocking pseudospin polarization.
For $E_{F} \to 0$ the conclusion is that the density-wave instability occurs before the anisotropy distortion, and that it will
occur for arbitrarily weak interactions \cite{min_MF,zhang_RG,levitov_MF}.

\section{Exciton Condensates in Double-Layer Graphene}
\begin{figure}[hbp]
\label{doublelayerfig}
\begin{centering}
\begin{tabular}{cc}
\quad \quad \quad \quad \quad \quad& \includegraphics[width=12cm,angle=0]{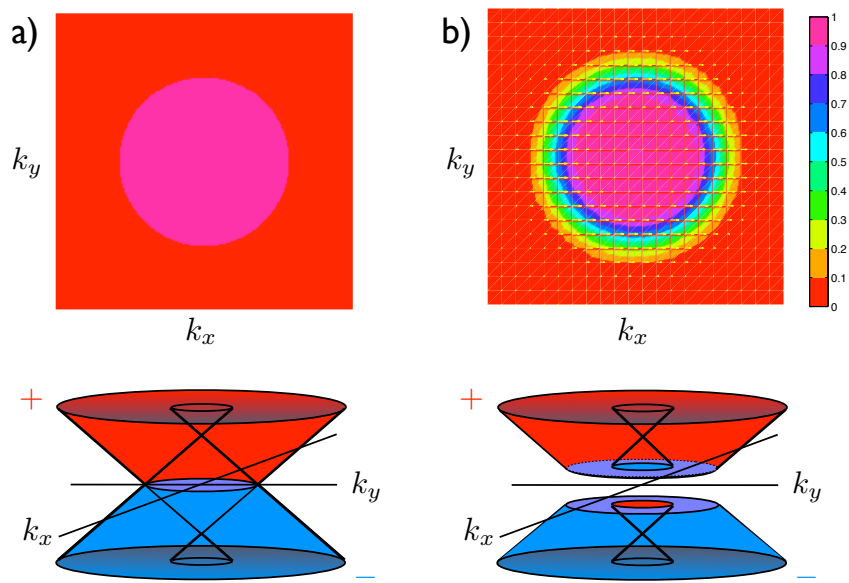}    \\
\end{tabular}
\end{centering}
\caption{
Pseudospin orientations and quasiparticle dispersions in double layer graphene near the Dirac point.
In the two-band pseudospin model the two bands furthest from the Fermi energy are not 
accounted for explicitly.
In double layer graphene tunneling between layers is 
negligible so that the pseudospin label is equivalent to a layer label.
The band pseudospin direction changes abruptly between up and down directions 
along the common Fermi surface where the conduction band of 
the high density layer and the valence band of the low density layer cross.
In presence of interactions a gap is opened  and 
pseudospin directions rotate gradually between up and down directions.
The figures (a) and (b) represent respectively the single particle
bands crossing at a Fermi circle, and the gapped phase with 
interlayer coherence. A gap opens 
in presence of arbitrarily weak interactions as in the case of bilayer graphene.
Any in plane pseudospin component introduces interlayer
coherence and reduces the total energy of the system. 
In our illustration we have chosen the interlayer coherence pseudospin component
to point in the $x$ direction.
}
\end{figure}
Now we turn to the case in which two graphene layers, one containing electrons in the conduction
band and the other containing an equal density of holes in the valence band, are coupled by repulsive Coulomb interactions.
If we ignore the completely full and completely empty energetically remote 
bands, we can view the double-layer graphene  system using the same pseudospin language
that we used for the monolayer and bilayer graphene cases.  In the absence of interactions the conduction band states of the
n-layer, which we associate with pseudospin up, are occupied inside a Fermi circle and the valence band states of the
p-layer, which we associate with pseudospin down, are occupied outside of this circle (see Fig. 4).  
As in the monolayer and bilayer cases, one pseudospin state is occupied at each momentum.   Because the band pseudospins are
oriented in the $\pm \hat{z}$ directions, rather than in the $\hat{x}-\hat{y}$ plane, the pseudospin distortions that
can potentially lower the energy are in the $\hat{x}$ and $\hat{y}$ direction distortions, rather than $\hat{\phi}$ and $\hat{z}$ distortions.
We find that
\begin{equation}
K_{\vec{k},\vec{p}}^{x,x}  = K_{\vec{k},\vec{p}}^{y,y} =
 \left[ 2A (\hbar v_{0} (k-k_{F}) +\Sigma_{\vec{k}}) \delta_{\vec{k},\vec{p}}
- V_{D}(\vec{k}-\vec{p})\right]\,,
\end{equation}
where
\begin{equation}
\Sigma_{\vec{k}} =  \frac{{n}_{{\rm b}\,\vec{k}}^{z}}{2A} \, \sum_{\vec{p}} V_{\rm S} (\vec{p}-\vec{k})\;
(1+ {n}_{{\rm b}\,\vec{p}}^{z}).
\end{equation}
(We take the band energy to be the quasiparticle energy in the absence of carriers in either layer and 
include a self-energy contribution from the full valence band of the n-type layer.) 
The sudden change in pseudospin orientation at the Fermi circle has a cost in
exchange energy which can be mitigated by rotating pseudospins into the $\hat{x}-\hat{y}$ plane,
which corresponds to establishing coherence between layers spontaneously.  Because $K^{x,x}=K^{y,y}$,
the pseudospin rotation can occur with the same gain in energy at any azimuthal angle.  The fact that the
energy is independent of the interlayer phase ({\em i.e}, the pseudospin azimuthal angle) implies that
this broken symmetry state supports super currents that flow in opposite directions in opposite layers\cite{min_exciton,su}.
As in the monolayer and bilayer cases, instability is indicated by a vanishing
$K_{\vec{k},\vec{p}}^{x,x}$ eigenvalue.  The instability condition can be found by solving
\begin{equation}
\frac{1}{A} \sum_{\vec{p}} \, K_{\vec{k},\vec{p}}^{x,x} \delta^{x}_{\vec{p}}  \equiv 0\,
\end{equation}
with $\delta^{x}_{\vec{p}} \to C/\hbar v_{0} |k-k_{F}|$.  For a $\delta$-function interaction the self-energy
term simply shifts the relationship between Fermi energy and density and plays no role.  The instability
criterion is therefore $ U \nu^{\rm DL} \ln(2 E_{F}/\delta) > 2$ where we have chosen $2 E_{F}$ as an
ultraviolet cutoff, $\delta$ is an infrared cut-off, and $\nu^{\rm DL}=E_{F}/(2\pi \hbar^2 v_{0}^2)$ is the constant density-of-states of
the double-layer Dirac model for energies between $0$ and $E_{F}$.  As in the bilayer case, an
instability occurs for arbitrarily weak interactions.  Although this conclusion is universally accepted
by researchers who have examined this possible ordered state\cite{min_exciton,joglekar,lozovik,efetov,han}
estimates of the size of the consequent energy gap vary widely because of the difficulty of accounting 
accurately for the influence of carrier screening.  

\section{Discussion}
Graphene two-dimensional electron systems (2DES) are remarkable for several different reasons.
The fact that they are truly two-dimensional on an atomic length scale elevates 2DES physics from the low-temperature
world to the room-temperature world.  
Furthermore, they are accurately described by very simple models over very wide
energy ranges and yet have electronic properties that can be qualitatively altered simply by stacking\cite{mccann_BLG,multi1,multi2,zhang_ABC,guinea_stack}  
them in different arrangements, and by adjusting external gate voltages.
In this article we have discussed the properties of three different 
graphene 2DES's using a simple mean-field-theory pseudospin language and specializing 
to the case of electrically neutral systems.

The basic building block of all graphene 2DES's is the isolated monolayer, which is described by a
massless Dirac $\vec{k}\cdot\vec{p}$ Hamiltonian over a wide energy range.
The Dirac model has chiral quasiparticles, and in the graphene case the chirality refers to the relationship
between $\vec{k}\cdot\vec{p}$ momentum and the direction of a pseudospin associated with the sublattice
degree-of-freedom of graphene's honeycomb lattice \cite{min_MF,maxim}.  In neutral graphene each momentum is singly occupied
on average.  In our mean-field-theory approach we do not account for quantum fluctuations in these
momentum occupation numbers, but allow energy to be minimized with respect to the pseudospin orientation
at each momentum.  Using this approach we find that in
monolayer graphene strong interactions can lead to a broken symmetry state in
which the pseudospin rotates from the $\hat{x}-\hat{y}$ plane toward the
$\pm \hat{z}$ directions, breaking inversion symmetry and opening up a gap in the quasiparticle
excitation spectrum.  
This conclusion has been reached previously\cite{previoussinglelayer,fujita,afqmc1,afqmc2,meng,mono_hf} 
by several researchers, sometimes using more sophisticated theoretical approaches which 
attempt quantitative estimates of the required interaction strengths. 
In $\vec{k}\cdot\vec{p}$ mean-field theory the broken symmetry state consists of a
density-wave state with more charge on one sublattice than the other within each spin-valley flavor, but
no overall charge density variation.  Since exchange interactions beyond the $\vec{k}\cdot\vec{p}$ level
favor\cite{jung_MF} states with the same sublattice polarization on both valleys, the strong interaction
state is likely a spin-density-wave state.  As it happens, it appears to be clear from experiment
that this broken symmetry state
does not occur\cite{geim,rotenberg} in monolayer graphene, even when suspended. 
This property is generally consistent with theoretical expectations and is perhaps
unfortunate from the point of view of researchers interested
in many-body phenomena.  Happily the very closely related bilayer and double-layer graphene systems that 
were not anticipated prior to the experimental emergence of the graphene field, 
are more likely to have broken symmetry states and may save the day for interesting many-body 
phenomena in graphene based 2DES's.

Bilayer graphene is described, over a more limited energy range however, by a similar pseudospin $\vec{k}\cdot\vec{p}$
model with quadratic rather than linear dispersion; in bilayer graphene the quasiparticle velocity vanishes with
momentum as in a conventional 2DES, but the pseudospin direction still depends on the direction of momentum.
In pseudospin language, it is clear that the bilayer is more susceptible to instabilities to broken symmetry
states because the cost in band energy of rotating the pseudospins toward the $\pm \hat{z}$ direction
is smaller at small momentum.  Indeed we find that instabilities occur for arbitrarily weak interactions\cite{min_MF,zhang_RG}.
In the bilayer case the spin-density-wave state has opposite pseudospin
orientations for opposite spins\cite{zhang_SQH,jung_MF,zhang_ZM}.   
An alternate state in which the pseudospin orientation is
rotated within the $\hat{x}-\hat{y}$ plane to increase the degree of alignment leads to an anisotropic state, and
gains less exchange energy for a given interaction energy cost.
There is indeed a great deal of evidence from recent experiments\cite{yacoby_LC,yacoby_SQH}
that a broken symmetry state does occur in bilayer graphene, but the character of the
state is not yet completely settled since some experimenters
find evidence for a 
gapless anisotropic state\cite{geim_nematic} and others\cite{lau_SQH,schonenberger_SQH}
evidence for a gapped isotropic state\cite{min_MF,zhang_RG,levitov_MF,jung_MF}.  
On the theoretical side different researchers have also reached
different conclusions concerning the character of these states, 
with some researchers\cite{vafek_RG,falko_RG}
concluding that the broken symmetry state should be anisotropic.  

In mean-field theory the gapped broken symmetry state has lower
energy than the anisotropic broken symmetry state because all band pseudospinors can be tilted to the $\pm \hat{z}$ 
direction to reduce the probability of finding electrons with parallel pseudospins.
The efficacy of in-plane pseudospin distortions is reduced because their pseudospin-directions $\hat{\phi}_{\vec{k}}$ 
depend on the momentum orientation angle $\phi_{\vec{k}}$.  
To us, the conclusion that the weak-coupling broken symmetry state will be gapped
appears to be unavoidable unless somehow overturned, in a way which has not yet been clearly articulated,
by inter-flavor correlations.  In drawing conclusions from the renormalization group 
calculations\cite{zhang_RG,vafek_RG,falko_RG} which attempt to go beyond the 
mean-field theory considerations described here, it is important to realize that,
because short-range interactions within a valley act only between opposite pseudospins,
the pseudospin dependence of the corresponding flowing interaction  
has no significance in a many-fermion Hilbert space.

Bilayer graphene differs from monolayer graphene mainly because of its weaker dependence of band energy on
pseudospin direction at small momentum.  This difference is sufficient to lead to states with
broken pseudospin symmetry.  
Double-layer graphene differs in a more qualitative way because
not only the interaction energy, but also the band energy, is diagonal in the $\hat{z}$ component of
pseudospin, {\em i.e.} in layer index.  The band Hamiltonian in this case has a sudden change in the
sign of the $\hat{z}$ direction pseudospin orientation.  This momentum space {\em domain wall} has a
large interaction energy cost which can be mitigated by rotating the pseudospins near the Fermi surface out
of the $\hat{z}$ direction with a common azimuthal angle.  In this way, the momentum dependence of the
pseudospin rotates smoothly between $\hat{z}$ and $-\hat{z}$ directions, and the interaction 
energy is lowered.
Like bilayer graphene, double-layer graphene has a broken symmetry for arbitrarily weak interactions.
In both cases the size of the gap is difficult to estimate quantitatively
\cite{min_exciton,joglekar,lozovik,efetov,han} and likely to be 
overestimated by mean-field-theory
The quasiparticle density-of-states remain finite at low energies in the
double-layer graphene case because the pseudospin field vanishes along a line in
momentum space while in the bilayer case because it varies quadratically with wavevector near $k=0$,
but the end result is essentially the same.

In two-dimensional electron systems, the quantized Hall conductance can be
expressed\cite{TKNN} in terms of an integral of Berry curvature over momentum space.
Using ideas from topology it is possible to show that the Hall conductivity is always
$e^2/h$ times an integer valued topological invariant known as Chern number.
For two-band models, momentum states can always be described using a pseudospin-$1/2$ language
like the one used in this paper.  For this case the quantized Hall conductivity is especially
simple to visualize geometrically since it is equal to the number of times the unit sphere of
pseudospin directions is covered upon integrating over momentum space.
The broken symmetry states of bilayer graphene carry a positive or negative unit
of Hall conductivity because they
cover either the north-pole or the south pole twice when the pseudospin at $\vec{k}=0$ points to the
north or south pole.  (In the double-layer graphene case, on the other hand, the pseudospin is confined to
the interface between a plane and the pseudospin unit sphere, {\em i.e.} to a line on the unit sphere that encompasses
zero area.)  The inversion symmetry breaking states in bilayer graphene can therefore be viewed as 
spontaneous quantum Hall states\cite{levitov_QAH, zhang_SQH, jung_MF}.  
In the spin-density-wave state the Hall contributions from different valleys cancel, 
but states with a non-zero Hall conductivity can be stabilized  
by going to a non-zero total carrier density in weak magnetic fields.

In this article we have discussed the possibility of interaction driven broken symmetries 
in three different graphene based two-dimensional electron systems, monolayer graphene 
systems and two layer graphene systems that are stacked in two different ways.  The 
three cases we have discussed are most easily addressed theoretically, but not by 
any means the ones that are most likely to have strong interaction effects and broken symmetries.
ABC stacked multilayers\cite{multi1,multi2,zhang_ABC}, for example, tend to have even smaller 
separations between conduction and valence bands at small $|\vec{k}|$, but are 
complicated by competing electronic structure details.   Recent advances in techniques
for preparing samples in which disorder plays an inessential role appear to be bringing 
us close to clear experimental conclusions as to the strength and character of broken symmetries in 
bilayers and trilayers.  These developments will, no doubt, reveal some surprises that present 
some focused challenges to theory.

\end{document}